\documentclass[aps,pre,reprint,amsmath,amssymb]{revtex4-2}
\usepackage{txfonts}
\usepackage{graphicx}
\usepackage{bm}

\begin{document}

\title{Efficient correlation-based discretization of continuous variables for annealing machines}

\author{Yuki Furue$^1$, Makiko Konoshima$^2$, Hirotaka Tamura$^3$, and Jun Ohkubo$^1$}

\affiliation{$^1$Graduate School of Science and Engineering, Saitama University, 255 Shimo-ohkubo, Sakura-ku, Saitama, 338--8570 Japan\\
  $^2$Fujitsu Limited, 4--1--1, Nakahara-ku, Kawasaki-shi, Kanagawa, 211--8588 Japan\\
$^3$DXR laboratory Inc., 4-38-10, Takata-Nishi, Kohoku-Ku, Yokohama-shi, Kanagawa, 223--0066 Japan} 

\begin{abstract}
Annealing machines specialized for combinatorial optimization problems have been developed, and some companies offer services to use those machines. Such specialized machines can only handle binary variables, and their input format is the quadratic unconstrained binary optimization (QUBO) formulation. Therefore, discretization is necessary to solve problems with continuous variables. However, there is a severe constraint on the number of binary variables with such machines. Although the simple binary expansion in the previous research requires many binary variables, we need to reduce the number of such variables in the QUBO formulation due to the constraint. We propose a discretization method that involves using correlations of continuous variables. We numerically show that the proposed method reduces the number of necessary binary variables in the QUBO formulation without a significant loss in prediction accuracy.
\end{abstract}

\maketitle

\section{Introduction}
\label{introduction}

Combinatorial optimization problems aim to find the optimal combination satisfying certain constraints. It is crucial to obtain combinatorial optimization problems quickly in practice. However, a computational explosion occurs when the number of combinations becomes enormous; a brute-force method cannot solve this problem.

Simulated and quantum annealing principles in physics play a crucial role in developing specialized machines for combinatorial optimization problems. Typical examples include D-Wave Advantage by D-Wave Systems Inc.\cite{d-wave}, Digital Annealer by Fujitsu Ltd.\cite{digital_annealer}, and CMOS Annealing Machine by Hitachi Ltd.\cite{cmos} The input form of these annealing machines is a quadratic form of binary variables called the quadratic unconstrained binary optimization (QUBO) formulation, which is equivalent to the Ising model. Hence, one should convert the original combinatorial optimization problems to the QUBO formulation. Then, the specialized machines quickly explore the optimal solution based on simulated or quantum annealing principles. Reference \cite{ising_formulation} gives various examples of the Ising models and QUBO formulations for famous combinatorial optimization problems.

Although specialized machines give only approximate solutions, we can use them in various practical situations. Many research institutes and companies are investigating applications of annealing machines to address social issues. Such applications include machine learning \cite{qloss, k-means, black-box}, financial optimization \cite{portfolio}, space debris removal \cite{space_debri}, traffic flow \cite{AGV, traffic}, medical imaging \cite{IMRT}, and E-commerce \cite{e-commerce}. As mentioned above, the annealing machines can solve combinatorial optimization problems that are difficult to solve with conventional computers within reasonable calculation time. Hence, they are attracting much attention as a new computational technology.

Annealing machines with the QUBO formulation can also solve problems with continuous variables; binary expansions of the original continuous variables are used \cite{encoding}. The corresponding QUBO formulation is then derived. There are studies on the binary expansion method with which annealing machines solve linear regression problems \cite{linear_regression} and support vector machines \cite{support_vector_machine}. Annealing machines reduce the computational time by a factor of 2.8 for relatively large linear regression problems \cite{linear_regression}; this study showed the superiority of annealing machines.

However, annealing machines have the limitation that the number of variables is not very large. Practical optimization problems require many variables, which makes it impossible to embed the QUBO formulation. For example, D-Wave Advantage can only solve problems up to about 5000 qubits. As mentioned above, continuous optimization problems require the discretization of variables. Previous studies used a naive discretization with a fixed number of binary variables for each continuous variable. Hence, the number of necessary binary variables increases proportionally with the number of continuous variables. It is problematic for a large-scale optimization problem with continuous variables. A rough discretization decreases the prediction accuracy because of insufficient resolution.

We propose an efficient discretization method for annealing machines, which involves the correlation between continuous variables. We carry out a short sampling for the original problems with continuous variables in advance. Then, the information on the correlations leads to a problem-specific discretization. The basic idea is as follows. We share the binary variables between the continuous variables that tend to be roughly the same value. Thus, the number of binary variables can be reduced without degrading the prediction accuracy. We investigated the effectiveness of the proposed method through a numerical experiment on linear regression problems.

The remainder of the paper is as follows. In Sec.~2 we give a brief review of the formulation for the annealing machines. We review the linear regression problem and its expression as a QUBO formulation in Sec.~3. In Sec.~4, we present the proposed efficient correlation-based discretization method. In Sec.~5, we explain the numerical experiment to investigate the effectiveness of the proposed method. Finally, we conclude the paper in Sec.~6.

\section{Basic Formulation for Annealing Machines}

\subsection{Ising model and QUBO formulation}

The Ising model is one of the most fundamental models to describe the properties of magnetic materials. Let $\sigma_i \in \{-1, 1\}$ denote a variable for the $i$-th spin. Then, the energy function of the Ising model with the state vector $\bm{\sigma}$ is defined as follows:
\begin{align}
  \label{eq:ising_model}
  E(\bm{\sigma}) = -\sum_{i < j} J_{ij} \sigma_i \sigma_j - \sum_i h_i \sigma_i,
\end{align}
where $J_{ij} \in \mathbb{R}$ corresponds to the two-body interaction between the spins $\sigma_i$ and $\sigma_j$, and $h_i \in \mathbb{R}$ is the external magnetic field on $\sigma_i$.

The QUBO formulation has the following cost function for the state vector $\bm{z}$:
\begin{align}
  E(\bm{z}) = -\sum_{i, j}Q_{ij}z_iz_j,
  \label{eq:qubo}
\end{align}
where $z_i \in \{0,1\}$ is $i$-th binary variable in $\bm{z}$, and $Q_{ij} \in \mathbb{R}$ is the strength of the interaction between the binary variables $z_i$ and $z_j$. This formulation is equivalent to the Ising model via the variable transformation with 
\begin{align}
  z_i = \frac{1 + \sigma_i}{2}.
  \label{eq:variable_change}
\end{align}
Conversion between $\{J_{ij}\}, \{h_i\}$ and $\{Q_{ij}\}$ is also possible.

From the computational viewpoint, annealing machines are domain-specific, and their role is simple; to find the ground state that minimizes Eqs.~\eqref{eq:ising_model} or \eqref{eq:qubo}. Despite the domain-specific characteristic, we can solve various combinatorial optimization problems with annealing machines. The reason is that various optimization problems are converted into the QUBO formulation \cite{ising_formulation}; we formulate the Ising model or QUBO formulation so that the ground state coincides with the optimal solution of the combinatorial optimization problem. Hence, we can solve combinatorial optimization problems with the annealing machines.

As stated above, the Ising model and the QUBO formulation are equivalent. However, the QUBO formulation is more suitable for computation, and then we employ the QUBO formulation below.

\subsection{Two types of annealing machines}
\label{seq:anenaling}

The QUBO formulation is widely applicable to different types of annealing machines.

There are two main types of annealing machines currently under development. One uses simulated annealing (SA) and the other uses quantum annealing (QA). Although the mechanisms of the search process differ, both types accept the QUBO formulation as the input for specific hardware. We briefly review why the QUBO formulation is suitable for both types of machines.

SA is an algorithm based on thermal fluctuations. When the temperature is high, the thermal fluctuations are large. The algorithm searches the large area in the state spaces. A gradual decrease in temperature indicates the settling down to the ground state. Numerically, the probability $P$ of a state change is often defined as follows:
\begin{align}
  P = \min\left[1, \exp\left(-\frac{\Delta E}{T}\right)\right],
  \label{eq:probability}
\end{align}
where $\Delta E$ is the energy difference when a certain state is changed, and $T$ is temperature. The ground state is adequately obtained if the temperature decreases slowly enough \cite{gibbs}. Note that $\Delta E$ in Eq.~\eqref{eq:probability} is evaluated using Eq.~\eqref{eq:qubo}. The algorithm uses the cost function of the QUBO formulation. It indicates that the QUBO formulation is suitable as the input for machines.

QA was proposed by Kadowaki and Nishimori and uses quantum fluctuation \cite{quantum_annealing}. Instead of the classical binary spin variables, we need quantum spins denoted by the Pauli matrix. Introducing the state vector with the $z$ component of the Pauli matrices, $\hat{\bm{\sigma}}^z$, and the $x$ component for $i$-th spin, $\hat{\sigma}_i^x$, the Hamiltonian is defined as follows:
\begin{align}
  \widehat{H} = E(\hat{\bm\sigma}^z)-\Gamma\sum_i \hat{\sigma}^x_i,
  \label{eq:qa_ising_model}
\end{align}
where $\Gamma$ is a control parameter. The second term in the right-hand side (r.h.s.) of Eq.~\eqref{eq:qa_ising_model} corresponds to the quantum fluctuations. The initial state is prepared as the ground state under quantum fluctuations. By gradually weakening the quantum effect, i.e., $\Gamma$, QA seeks the ground state determined by the first term on the r.h.s. of Eq.~\eqref{eq:qa_ising_model}. QA can use more rapid scheduling than SA \cite{quantum_annealing_scheduling}. Such a computation method is also called adiabatic quadratic computing \cite{adiabatic}.

The annealing mechanism of QA differs from that of SA. However, the final ground state is defined only by the first term on the r.h.s. of Eq.~\eqref{eq:qa_ising_model}. It is straightforward to obtain the first term from the QUBO formulation in Eq.~\eqref{eq:qubo}. Hence, practical QA machines, such as the D-Wave Advantage, accept the QUBO formulation as the input. 

The main topic of this paper is the discretization of continuous variables. Although the annealing mechanisms vary between SA and QA, we can use the following discussions with the QUBO formulation for both SA and QA machines. Because of the simpler notations, only the classical cases, i.e., SA cases, are considered in the following sections; we do not consider any quantum effects.

\section{Linear Regression and QUBO Formulation}

\subsection{Linear regression problems}

Linear regression is one of the fundamental methods in data analysis with which we seek the relationship between a response variable and explanatory variables. Although matrix operations are sufficient to solve simple linear regression problems, further computational efforts are necessary in large-scale cases or when there are regularization terms. Hence, the use of an annealing machine is desirable. With these problems in mind, Date and Potok discussed the implementation and increasing the speed of simple linear regression problems using an annealing machine \cite{linear_regression}. Linear regression problems also yield an example to investigate discretization methods of continuous variables. We briefly review Date and Potok's discussion.

A linear regression model is defined as
\begin{align}
  y = w_1 + w_2 x_{2} + \cdots + w_D x_{D},
\label{eq_regression_model}
\end{align}
where $w_d \in \mathbb{R} \,\, (d = 1, \dots, D)$ is the $d$-th parameter, and $x_{d} \in \mathbb{R} \,\, (d = 2, \dots, D)$ is the $d$-th input variable. They are summarized in the following forms:
\begin{align}
\bm{w} &= [w_1, w_2, \cdots, w_D]^\mathrm{T} \in \mathbb{R}^{D}, \\
\bm{x} &= [1, x_{2}, \cdots, x_{D}]^\mathrm{T} \in \mathbb{R}^{D},
\end{align}
where the first component in $\bm{x}$ is a dummy variable; the dummy variable is convenient for the vector expression of Eq.~\eqref{eq_regression_model} as follows:
\begin{align}
  y = \bm{w}^\mathrm{T} \bm{x}.
\end{align}

In linear regression problems, we seek the model parameters $\bm{w}$ in Eq.~\eqref{eq_regression_model} that are suitable for our obtained data. Let $\{(\bm{x}_i,y_i):i=1,2,\cdots,N\}$ be a set of $N$ training data, where $\bm{x}_i = [1, x_{i2}, \cdots, x_{iD}]^\mathrm{T} \in \mathbb{R}^{D}$ is a $D$-dimensional input vector, and $y_i \in \mathbb{R}$ is an output for $i$-th data. We introduce an output vector $\bm{y} = [y_1, y_2, \cdots, y_N]^\mathrm{T}$ and matrix $X \in \mathbb{R}^{N \times D}$ defined as
\begin{align}
X = \begin{bmatrix}
    \, 1 & x_{12} & \cdots & x_{1D} \\
    \, 1 & x_{22} & \cdots & x_{2D} \\
    \, \vdots & \vdots & \ddots & \vdots \\
    \, 1 & x_{N2} & \cdots & x_{ND} \\
\end{bmatrix}.
\end{align}
Then, a conventional linear regression problem has the following cost function:
\begin{align}
  \widetilde{E}(\bm{w}) = \|\bm{y}-X\bm{w}\|^2=\bm{y}^\mathrm{T}\bm{y}-2\bm{w}^\mathrm{T}X^\mathrm{T}\bm{y}+\bm{w}^\mathrm{T}X^\mathrm{T}X\bm{w}.
  \label{eq:pre_reg_cost}
\end{align}
In Eq.~$\eqref{eq:pre_reg_cost}$, $\bm{y}^\mathrm{T}\bm{y}$ is meaningless in optimizing $\bm{w}$. Hence, Eq.\:$\eqref{eq:pre_reg_cost}$ comes down to the following problem:
\begin{align}
  E(\bm{w}) = \bm{w}^\mathrm{T}X^\mathrm{T}X\bm{w}-2\bm{w}^\mathrm{T}X^\mathrm{T}\bm{y}.
  \label{eq:reg_cost}
\end{align}
Of course, it is possible to find the minimum for Eq.~\eqref{eq:reg_cost} via simple matrix operations. However, the matrix operations include a matrix inversion with a high computational cost. Annealing machines directly solve the optimization problem in the form of Eq.~\eqref{eq:reg_cost}.

\subsection{QUBO formulation for linear regression}
\label{sec:pre_research}

As denoted in Sec.~\ref{introduction}, Date and Potok showed that the QUBO formulation and the annealing machines enable faster computation than conventional classical computers even in simple linear regression problems without any regularization term \cite{linear_regression}. We now explain the QUBO formulation for simple linear regression problems.

First, we introduce $\bm{b} = [b_1, b_2, ..., b_K]^\mathrm{T} \,\, (K \in \mathbb{N})$ as a basis vector. For later use, the components in $\bm{b}$ are in ascending order of the absolute value. An example of this vector is $\bm{b}=[\frac{1}{2}, -\frac{1}{2}, 1, -1, 2, -2]^\mathrm{T}$. 

Second, we introduce a $D \times K$ binary matrix $\widetilde{\bm{z}}\in\{0, 1\}^{D \times K}$. Then, the discretization of the linear regression weight $w_i$ is 
\begin{align}
  w_i = \sum_{k=1}^{K}b_k\widetilde{z}_{ik} \quad (i \in 1,2,\dots,D),
  \label{eq:w_i}
\end{align}
which indicates that the binary variable $\widetilde{z}_{ik}$ acts like a flag to determine whether the basis $b_k$ is used for the expression. For the above example of $\bm{b}=[\frac{1}{2}, -\frac{1}{2}, 1, -1, 2, -2]^\mathrm{T}$, we can express the values with $\{-\frac{7}{2}, -3, \dots, 3, \frac{7}{2}\}$. Note that there may not be a one-to-one correspondence between $w_i$ and $\widetilde{\bm{z}}$; we can use redundant expressions for original variables. Although the redundant basis could yield good annealing performance, this point is not our main research topic.

Third, let $\bm{z}=[\widetilde{z}_{11},\cdots,\widetilde{z}_{1K},\widetilde{z}_{21},\cdots,\widetilde{z}_{2K},\cdots,\widetilde{z}_{D1},\cdots,\widetilde{z}_{DK}]^T$ denote the flattened one-dimensional vector derived from the binary matrix $\widetilde{\bm{z}}$. Then, we rewrite Eq.~\eqref{eq:w_i} in a matrix form as follows:
\begin{align}
  \bm{w} = B\bm{z},
  \label{eq:w}
\end{align}
where $B = I_D \otimes \bm{b}$ is the basis matrix derived from the Kronecker product of the $D$-dimensional unit matrix $I_D$ and $\bm{b}$. An example of the matrix $B$ is given later.

Finally, we obtain the QUBO formulation of linear regression from Eqs.~\eqref{eq:reg_cost} and \eqref{eq:w} as follows:
\begin{align}
  E(\bm{z}) = \bm{z}^\mathrm{T}B^\mathrm{T}X^\mathrm{T}XB\bm{z}-2\bm{z}^\mathrm{T}B^\mathrm{T}X^\mathrm{T}\bm{y}.
  \label{eq:lr_qubo}
\end{align}

\section{Proposed Efficient Correlation-based Discretization Method}
\label{sec:bit_cut_method}

\subsection{Problems with naive method}
\label{sec:problems_of_naive_method}

As mentioned in Sec.~\ref{sec:pre_research}, the size of the binary vector $\bm{z}$ is $D \times K$ because we assign $K$ binary variables to each continuous variable with the naive discretization method. However, for large-scale problems, this method is likely to exceed the size limit of the annealing machine. How can we reduce the number of binary variables within the size limit of the annealing machine?

The simplest method is to reduce the components of the basis assigned in the binary expansion. However, this discretization reduces the precision of the original continuous variables and the ranges that the variables can represent. 

One may employ the discretization method for each continuous variable. The flexible discretization may yield good performance. However, we cannot determine the required expression for each original continuous variable before solving the problem. 

It is also possible to choose randomly continuous variables with reduced bases; this randomized method might yield a good accuracy on average. However, it could degrade the accuracy of a realization.

In summary, the reduction in the basis degrades the precision of each original continuous variable and yields narrow ranges. Hence, this reduction could lead to poor prediction accuracy for the linear regression.

\subsection{Basic idea of proposed method}

To avoid the degradation in precision and narrower ranges, we exploit the same binary variables for discretization. For example, $\bm{b}=[\frac{1}{2}, -\frac{1}{2}, 1, -1, 2, -2]^\mathrm{T}$ expresses $-2.5$ and $-1.5$ as
\begin{align*}
-2.5 &= \begin{bmatrix} 0,1,0,0,0,1 \end{bmatrix} \bm{b} \\
&= 0 \times b_1 + 1 \times b_1 + 0 \times b_2 + \cdots + 1 \times b_6, \\
-1.5 &= \begin{bmatrix} 1,0,0,0,0,1 \end{bmatrix} \bm{b} \\
&= 1 \times b_1 + 0 \times b_1 + 0 \times b_2 + \cdots + 1 \times b_6. 
\end{align*}
When we consider two large negative values, $b_6$ will commonly be used. Hence, the coefficient of $b_6$ is always $1$. If two variables are strongly correlated, the coefficients of $b_6$ for them will be the same. That is, the change in the coefficients of $b_6$ largely affects these two variables, which indicates the strong correlation between them. Hence, we can use the same binary variable for $b_6$ in this example.

Note that using the same binary variable does not degrade the precision of approximation, and the range is unchanged. These are different from the naive methods in Sec.~\ref{sec:problems_of_naive_method}. Note that we must find pairs of continuous variables with strong correlations. Hence, we need a preliminary evaluation of correlations in the original continuous variables. Although this evaluation takes some computational effort, only rough estimations for the correlations are sufficient; the computational cost is not high.

The basic idea of the proposed method is summarized as follows: evaluate correlations, find strongly correlated pairs, and reduce redundant binary variables. In the following sections, we first explain how to reduce the number of binary variables and then give the entire procedure to derive the QUBO formulation.

\subsection{Choice of basis matrix}
\label{sec:choice_of_basis_matrix}

As reviewed in Sec.~\ref{sec:pre_research}, continuous parameters $\bm{w}$ are discretized using Eq.~\eqref{eq:w}. For example, we here consider $D=2$, $K=3$, $\bm{b} = [b_1, b_2, b_3]^\mathrm{T}$. Then, Eq.~\eqref{eq:w} yields the following discretization:
\begin{align}
\begin{bmatrix}
    w_{1} \\
    w_{2} \\
\end{bmatrix}
=
\begin{bmatrix}
    b_{1} & b_{2} & b_{3} & 0 & 0 & 0\\
    0 & 0 & 0 & b_{1} & b_{2} & b_{3}\\
\end{bmatrix}
\begin{bmatrix}
    \widetilde{z}_{11} \\
    \widetilde{z}_{12} \\
    \widetilde{z}_{13} \\
    \widetilde{z}_{21} \\
    \widetilde{z}_{22} \\
    \widetilde{z}_{23} \\
\end{bmatrix}.
\label{eq:matrix_w_example}
\end{align}
That is, $3$ bits are assigned to each of the continuous variables $w_1$ and $w_2$, and the total number of binary variables is $6$.

We assume that the continuous variables $w_1$ and $w_2$ are strongly correlated. We then modify matrix $B$ by considering the strong correlation between the continuous variables. In accordance with the modification, the size of $\bm{z}$ is also reduced. The result of the modification is as follows:
\begin{align}
\begin{bmatrix}
    w_{1} \\
    w_{2} \\
\end{bmatrix}
  =
\begin{bmatrix}
    b_{1} & b_{2} & 0 & 0 & b_{3}\\
    0 & 0 & b_{1} & b_{2} & b_{3}\\
\end{bmatrix}
\begin{bmatrix}
    \widetilde{z}_{11} \\
    \widetilde{z}_{12} \\
    \widetilde{z}_{21} \\
    \widetilde{z}_{22} \\
    \widetilde{z}_{23} \\
\end{bmatrix}.
\label{eq:bit_cut_example}
\end{align}
Note that the redundant binary variable $\widetilde{z}_{13}$ is deleted by sharing the binary variable $\widetilde{z}_{23}$ with $w_1$ and $w_2$. If further reduction is necessary, we can use the following expression:
\begin{align}
\begin{bmatrix}
    w_{1} \\
    w_{2} \\
\end{bmatrix}
  =
\begin{bmatrix}
    b_{1} & 0 & b_{2} & b_{3}\\
    0 & b_{1} & b_{2} & b_{3}\\
\end{bmatrix}
\begin{bmatrix}
    \widetilde{z}_{11} \\
    \widetilde{z}_{21} \\
    \widetilde{z}_{22} \\
    \widetilde{z}_{23} \\
\end{bmatrix}.
\label{eq:bit_cut_example2}
\end{align}

In summary, we seek a reduced matrix $B'$ and the corresponding vector for the binary variables $\bm{z}'$ with the following form:
\begin{align}
  \bm{w} = B'\bm{z}'.
  \label{eq:bitcut_w}
\end{align}
Note that the binary variable with a large absolute value is shared first. As mentioned in Sec.~\ref{sec:pre_research}, $\bm{b}$ is in ascending order. Therefore, the order of sharing is $b_K, b_{K-1}, \dots, b_1$. For a linear regression problem, there is a change in the optimization variables, and Eq.~\eqref{eq:lr_qubo} is modified as
\begin{align}
E(\bm{z}') = (\bm{z}')^\mathrm{T} (B')^\mathrm{T} X^\mathrm{T} X (B') \bm{z}' - 2 (\bm{z}')^\mathrm{T} (B')^\mathrm{T} X^\mathrm{T}\bm{y}.
\label{eq:lr_qubo_modified}
\end{align}
Note that the cost function is still in the QUBO formulation.

\subsection{Procedure to obtain reduced QUBO formulation}

To reduce the redundant binary variables, we should find strongly correlated pairs in advance. Hence, we carry out a short sampling procedure for the original problem with the continuous variables with a conventional computer. We then choose correlated pairs that share the same binary variables. We determine the correlated pairs using the correlation matrix with a certain threshold. 

Note that the threshold is determined by how many pairs we want to make; the limitation of the number of spins in the annealing machines affects this.

The procedure is summarized as follows:
\begin{enumerate}
  \item[(i)] Sample short-time series data of continuous variables using Monte Carlo methods.
  \item[(ii)] Calculate the correlations between continuous variables from the time series data.
  \item[(iii)] Define a basis matrix $B'$ and binary vector $\bm{z}'$ using the information on the correlations. That is, reduce matrix $B$ and $\bm{z}$ for the variable pairs whose correlation values exceed a certain threshold.
  \item[(iv)] Derive the QUBO formulation by using matrix $B'$ and the reduced $\bm{z}'$.
\end{enumerate}

We give some comments on this procedure. 

First, we need additional samplings in step (i). If the sampling step takes a long time, this procedure could not be practical even if the optimization is fast in annealing machines. However, we do not need highly accurate information on the correlations; only a rough estimation is sufficient for the pairing in step (ii). Hence, the short-time sampling is sufficient, and its computational cost is low. 

Second, one may obtain the correlations analytically in some cases, for example, in Gaussian random Markov fields. In such cases, we can skip the Monte Carlo samplings in step (i). 

Third, the number of reduced binary variables depends on the threshold with which we define the strongly correlated pairs. As stated above, we simply set the threshold in accordance with how much we want to reduce the number of variables.

\section{Numerical Experiment}
\label{sec:experiment}

As mentioned above, continuous variables with strong correlations tend to have the same value. Hence, the proposed method reduces the number of binary variables without a significant loss in prediction accuracy.

To verify the performance of the proposed method, we applied it to a linear regression problem. We calculate the mean absolute errors by generating artificial data. We compared the performance of the proposed method with that of the random reduction methods.

\subsection{Artificial data generation}

We use the following linear function to generate artificial data:
\begin{align}
  f(\bm{x}) =& 15.5+15.5x_1+10.0x_2+10.0x_3+5.0x_4+5.0x_5 \nonumber \\
  & -0.5x_6-0.5x_7-15.5x_8-15.5x_9.
\label{eq:target_problem}
\end{align}
Hence, the number of parameters is $D = 10$. First, we generate $x_d \, (d = 1, \dots, 9)$ from a uniform distribution $[-1,1]$ and we set it as the $i$-th input data $\bm{x}_i$. Observation noise $\eta_i \sim \mathcal{N}(0,1)$ is then sampled from the standard normal distribution. By adding $\eta_i$ to the true output $f(\bm{x}_i)$ as
\begin{align}
   y_i = f(\bm{x}_i) + \eta_i,
\end{align}
we have the $i$-th data $(\bm{x}_i, y_i)$.

We repeat this process $1000$ times and split the generated data into two sets, i.e., training and test. The size of the training data set is $100$, and the remaining $900$ data pairs are used as the test data set. Since the problem is not difficult to solve, we use a small amount of the training data.

\subsection{Samplings and evaluating the correlations}

Once we fix the training data, the cost function in Eq.~\eqref{eq:reg_cost} is determined. As the preparation step, we then obtain short-time series data from the Monte Carlo sampling for the linear regression cost function in Eq.~\eqref{eq:reg_cost}, which has the original continuous variables. Samples are generated from the Gibbs-Boltzmann distribution with the aid of the Metropolis rule. We generate the candidates for the Metropolis rule by adding a random variable generated from $\mathcal{N}(0,0.5)$ to a randomly selected parameter. The sampling interval is $2 \times D = 20$ steps, and the length of the time-series data is $100$. The sampling temperature is $0.1$, although we confirmed that the other temperatures are also suitable to obtain a rough estimation of the correlations.

\begin{figure}[b]
  \centering
  \includegraphics[keepaspectratio,width=90mm]{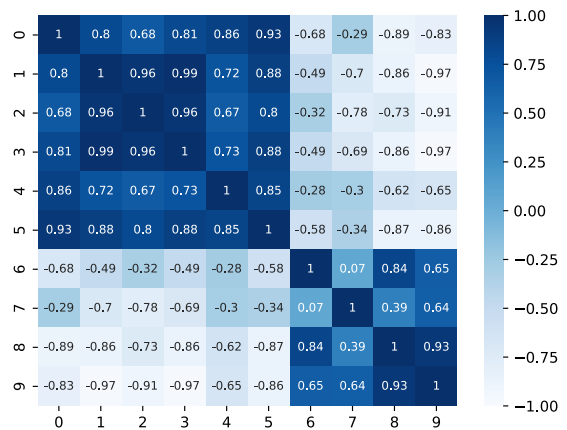}
  \caption{(Color online) Correlations between continuous variables. Short-time series data from the Monte Carlo samplings yields the correlations. Rough estimation is sufficient.}
  \label{fig:correlation}
\end{figure}

Figure~\ref{fig:correlation} shows an example of correlation matrices. There are several correlated pairs. The highest correlated one is $(w_1, w_3)$. Although $(w_1,w_2)$ and $(w_2,w_3)$ are candidates, we do not use them because $w_1$ and $w_3$ appear in the pair $(w_1, w_3)$. We then select the next pairs $(w_0, w_5)$ and $(w_8, w_9)$. We set the threshold to $0.8$. Hence, we reduced the binary variables from these three pairs, i.e., $(w_1,w_2)$, $(w_0, w_5)$, and $(w_8, w_9)$ for the case in Fig.~\ref{fig:correlation}.

\subsection{Creating basis matrix and binary vector}

Although we do not know detailed information about parameters $\{w_i\}$ in advance, it is necessary to define a basis vector in Eq.~\eqref{eq:w_i}. Since our aim was to show the effectiveness of the proposed method, we choose the basis vector $\bm{b}$ as follows:
\begin{align}
  \bm{b} =
  \begin{bmatrix}
    0.5 & -0.5 & 1 & -1 & 2 & -2 & 4 & -4 & 8 & -8
  \end{bmatrix}^{\mathrm{T}}.
  \label{eq:b}
\end{align}
Hence, $K=10$ in Eq.~\eqref{eq:w_i}. Note that the choice of the basis vector $\bm{b}$ is sufficient to represent the target function $f(\bm{x})$ in Eq.~\eqref{eq:target_problem}. The $\bm{b}$ yields the basis matrix $B$ via $B=I_D \otimes \bm{b}$.

Next, we redefine matrix $B'$ with the aid of the above-mentioned highly correlated pairs. For the selected pairs, we share certain binary variables, as exemplified in Sec.~\ref{sec:choice_of_basis_matrix}; the order is from highest absolute value to lowest. In the experiment, we changed the number of shared binary variables to each pair and investigated the decrease in the prediction accuracy.

Finally, we redefine $\bm{z}'$ in accordance with the redefined matrix $B'$. We then obtain the final QUBO formulation in Eq.~\eqref{eq:lr_qubo_modified} for the annealing machines.

\subsection{Annealing procedure}

We use a simple SA algorithm to find optimal solutions $\bm{z}'$ of the derived QUBO formulation. The SA simulations ran on a MacBookAir with Inter Core i5. In SA, the temperature schedule was as follows:
\begin{align}
T_t = T_0 \times \gamma^{t},
\end{align}
where $T_t$ is the temperature at the $t$-th iteration, $T_0$ is the initial temperature, and $\gamma$ is the decay rate. One iteration is defined as $2 \times \textrm{(\# of spins)}$ updates in accordance with the conventional Metropolis rule. After the iterations, we output the state that minimizes the QUBO formulation during the search procedures. Table~\ref{tb:annealing_parameter} shows the SA parameters.

The obtained $\bm{z}'$ leads to the final estimation for the continuous variables $\{ w_i \}$ via Eq.~\eqref{eq:bitcut_w}.

\begin{table}[b]
  \centering
  \caption{Parameters for annealing procedure}
  \begin{tabular}{l|c} \hline\hline
    Number of iterations & $10^6$ \\ 
    Initial temperature $T_0$ & $500$ \\ 
    Decay rate $\gamma$ for temperature scheduling & $0.99996$ \\ \hline\hline
  \end{tabular}
  \label{tb:annealing_parameter}
\end{table}

\subsection{Numerical results}

\subsubsection{Performance in prediction accuracy}

\begin{figure}[t]
  \includegraphics[keepaspectratio, width=90mm]{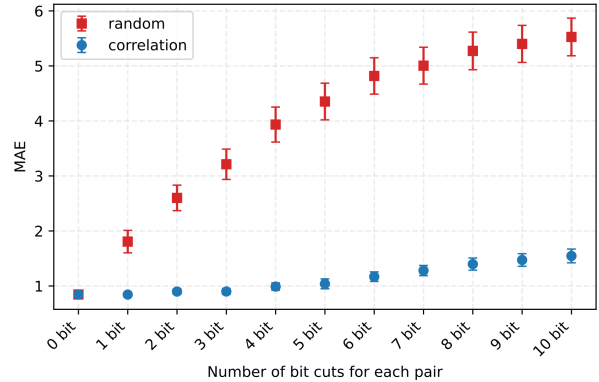}
  \caption{(Color online) Mean absolute error of annealing results. Filled squares show errors by employing the reduction method using randomly selected pairs. Filled circles correspond to those with the proposed method.}
  \label{fig:regression_compare}
\end{figure}

First, we investigated how the number of reductions affects the prediction accuracy. We used another reduction method for comparison, with which we randomly select the same number of pairs of continuous variables. We conducted $10$-fold cross-validation and evaluated the mean absolute errors (MAEs) and the standard deviations. Note that the correlation matrix and selected pairs vary with the training data set. Hence, the number of selected pairs varied with each training data set.

Figure~\ref{fig:regression_compare} shows the numerical results. The horizontal axis represents the number of reduced binary variables for each pair; for example, `2bit cut' indicates that a selected pair of continuous variables share two binary variables. Note that the total number of reduced binary variables was generally larger than the value of the horizontal axis because we selected several pairs of variables. The filled squares show the errors with the reduction method using randomly selected pairs. The filled circles correspond to those with the proposed method. The error bars indicate the standard deviations.

Since we added the noise with the standard normal distribution, the minimum value of MAE was about $1$. Figure~\ref{fig:regression_compare}  shows that the random reduction method performed worse due to increasing the number of shared bits. Even the $1$-bit cut resulted in a sudden decrease in the prediction accuracy.

We see a moderate increase in the MAE for the proposed, indicating its validity. Even the $10$-bit cut case yielded better accuracy than the $1$-bit cut case with the random reduction method, although this extreme case is not practical. 

Therefore, it is possible to reduce several shared binary variables without significant degradation in prediction accuracy.

\subsubsection{Performance in variable reduction and computational time}

\begin{table}[tb]
  \centering
  \caption{Total number of binary variables and computation time}
  \begin{tabular}{c|cc} \hline\hline
\begin{minipage}{20mm}
\begin{center}
Number of \\reduced binary variables per pair 
\end{center}
\end{minipage}
&
\begin{minipage}{25mm}
\begin{center}
Total number of\\ binary variables\\ (average for $10$ trials)
\end{center}
\end{minipage} 
&
\begin{minipage}{25mm}
\begin{center}
Computational time [sec]\\ (average for $10$ trials)
\end{center}
\end{minipage} \\
\hline
$0$ & $100 \pm 0$  & $331.7 \pm 1.7$\\
$1$ & $96.5 \pm 0.7$ & $309.0 \pm 4.5$\\
$2$ & $93.0 \pm 1.3$ & $287.4 \pm 8.0$\\
$3$ & $89.5 \pm 2.0$ & $266.6 \pm 10.9$\\
$4$ & $86.0 \pm 2.7$ & $247.2 \pm 14.6$\\
$5$ & $82.5 \pm 3.4$ & $228.7 \pm 17.2$\\
$6$ & $79.0 \pm 4.0$ & $210.2 \pm 20.0$\\
$7$ & $75.5 \pm 4.7$ & $193.0 \pm 22.4$\\
$8$ & $72.0 \pm 5.4$ & $176.2 \pm 24.3$\\
$9$ & $68.5 \pm 6.0$ & $160.2 \pm 26.3$\\
$10$ & $65 \pm 6.7$  & $145.3 \pm 26.8$\\
\hline\hline
  \end{tabular}
  \label{tb:number_of_bit}
\end{table}

Table~\ref{tb:number_of_bit} shows the number of binary variables and computational time; the values in the table are the average for ten trials in the cross-validation procedure. The first row in the table corresponds to the naive application of the QUBO formulation without any reduction.

We see that the reduction with $6$ variables per pair yielded about a 20\% reduction in the total number of variables. Even in the $6$-bit cut case, the prediction accuracy did not degrade much, as we see in Fig.~\ref{fig:regression_compare}. As stated in Sec.~\ref{introduction}, the limitation of available variables is severe with annealing machines. From a practical viewpoint, the proposed method will be crucial to embedding the QUBO formulation into the annealing machines. The reduction in variables also contributed to rapid computation, as shown in Table~\ref{tb:number_of_bit}.

\section{Conclusion}

We proposed an efficient correlation-based discretization method for annealing machines. The numerical experiment showed that it is possible to reduce the number of binary variables without a large loss in prediction accuracy. Recall that there are the following two constraints with annealing machines:
\begin{itemize}
  \item[1.] We need the QUBO formulation as the input; i.e., only binary variables are acceptable.
  \item[2.] There is a severe limitation on the number of binary variables; the number of binary variables for annealing machines is not large.
\end{itemize}
Compared with naive methods used in previous studies, the proposed method adequately addresses these two constraints. Note that we can flexibly increase or decrease the total number of binary variables by changing the correlation threshold. When one wants to use annealing machines for large-scale problems, the proposed method could be of great help in practice.

This study was the first attempt at a correlation-based reduction idea. Although the proposed method was verified, further studies are needed with other examples and more practical ideas. There are remaining issues. First, the proposed method handles only correlated pairs. There are various other ways to combine correlated variables. We will investigate which type of combination yields good performance. Second, we assumed the fully connecting type annealing machines. Some annealing machines are loosely coupled, such as the D-Wave Advantage and CMOS Annealing Machine. We need another procedure to embed the derived QUBO formulation into such machines.


\begin{thebibliography}{30}
  \bibitem{d-wave}
C. McGeoch and P. Farré, D-Wave Tech. Report Series 14-1049A-A (2020).

  \bibitem{digital_annealer}
M. Aramon, G. Rosenberg, E. Valiante, T. Miyazawa, H. Tamura, and H. G. Katzgraber, Front. Phys. \textbf{7}, 48 (2019).

  \bibitem{cmos}
T. Takemoto, K. Yamamoto, C. Yoshimura, M. Hayashi, M. Tada, H. Saito, M. Mashimo, and M. Yamaoka, IEEE Int. Solid-State Circ. Conf. \textbf{64}, 64 (2021).

  \bibitem{ising_formulation}
A. Lucas, Front. Phys. \textbf{2}, 2 (2014).

  \bibitem{qloss}
  V. S. Denchev, N. Ding, S. V. N. Vishwanathan, and H. Neven, Proc. 29th Int. Conf. Mach. Learn., 1003 (2012).

  \bibitem{black-box}
S. Izawa, K. Kitai, S. Tanaka, R. Tamura, and K. Tsuda, Phys. Rev. Res. \textbf{4}, 023062 (2022).

  \bibitem{k-means}
A. Davis and D. Prasanna, Quantum Inf. Process. \textbf{20}, 294 (2021).

  \bibitem{portfolio}
G. Rosenberg, P. Haghnegahdar, P. Goddard, P. Carr, K. Wu, and M. L. de Prado, IEEE J. Selected Topics in Signal Process. \textbf{10}, 1053 (2016).

  \bibitem{space_debri}
D. Snelling, E. Devereux, N. Payne, M. Nuckley, G. Viavattene, M. Ceriotti, S. Workes, G. di Mauro, and H. Brettle, Proc. 8th Eur. Conf. Space Debris (online) (2021).

  \bibitem{AGV}
M. Ohzeki, A. Miki, M. J. Miyama, and M. Terabe, Front. Comput. Sci. \textbf{1}, 9 (2019).

  \bibitem{traffic}
F. Neukart, G. Compostella, C. Seidel, D. von Dollen, S. Yarkoni, and B. Parney, Front. ICT \textbf{4}, 29 (2017).

  \bibitem{IMRT}
D. P. Nazareth and J. Spaans, Phys. Med. Biol. \textbf{60}, 4137 (2015).

  \bibitem{e-commerce}
N. Nishimura, K. Tanahashi, K. Suganuma, M. J. Miyama, and M. Ohzeki, Front. Comput. Sci. \textbf{1}, 2 (2019).

  \bibitem{encoding}
K. Tamura, T. Shirai, H. Katsura, S. Tanaka, and N. Togawa, IEEE Access \textbf{9}, 81032 (2021).

  \bibitem{linear_regression}
P. Date and T. Potok, Sci. Rep. \textbf{11}, 21905 (2021).

  \bibitem{support_vector_machine}
D. Willsch, M. Willsch, H. de Raedt, and K. Michielsen, Comput. Phys. Commun. \textbf{248}, 107006 (2020).


  \bibitem{gibbs}
S. Geman and D. Geman, IEEE Trans. Pattern Anal. Mach. Intell.  \textbf{PAMI-6}, 721 (1984).

  \bibitem{quantum_annealing}
T. Kadowaki and H. Nishimori, Phys. Rev. E \textbf{58}, 5355 (1998).

  \bibitem{quantum_annealing_scheduling}
S. Suzuki and M. Okada, J. Phys. Soc. Jpn. \textbf{74}, 1649 (2005).

  \bibitem{adiabatic}
E. Farhi, J. Goldstone, S. Gutman, J. Lapan, A. Lundgren, and D. Preda, Science \textbf{292}, 472 (2001).

\end{thebibliography}
\end{document}